\documentclass[aps,prx,longbibliography,twocolumn]{revtex4-1}
\usepackage[english]{babel}
\usepackage{array,amsmath,amsfonts,amssymb,dsfont,tabularx,multirow}
\usepackage{hhline}
\usepackage[T1]{fontenc} \usepackage{ae} \usepackage{color} 
\usepackage{amsmath}
\usepackage{graphicx}
\usepackage{soul}
\usepackage{siunitx}
\usepackage[dvipsnames]{xcolor}

\usepackage{textcomp, gensymb}

\newcommand{\LGM}{$\Lambda\text{GM}$}
\newcommand{\DLGM}{$\Delta_{\Lambda\text{GM}}~$}
\newcommand{\DLGMM}{\Delta_{\Lambda\text{GM}}}
\newcommand{\ket}[1]{|#1\rangle}

\usepackage{makecell}

\begin{document}


\title{Grey-molasses optical-tweezer loading: \\ Controlling collisions for scaling atom-array assembly}

\author{M. O. Brown}\thanks{These authors contributed equally}
\author{T. Thiele}\thanks{These authors contributed equally} 
\author{C. Kiehl}
\author{T.-W. Hsu}
\author{C. A. Regal}\email{regal@colorado.edu}
\affiliation{JILA, National Institute of Standards and Technology and University of Colorado, and
Department of Physics, University of Colorado, Boulder, Colorado 80309, USA}

\begin{abstract}

To isolate individual neutral atoms in microtraps, experimenters have long harnessed molecular photoassociation to make atom distributions sub-poissonian. While a variety of approaches have used a combination of attractive (red-detuned) and repulsive (blue-detuned) molecular states, to-date all experiments have been predicated on red-detuned cooling.  In our work, we present a shifted perspective -- namely, the efficient way to capture single atoms is to eliminate red-detuned light in the loading stage, and use blue-detuned light that both cools the atoms and precisely controls trap loss through the amount of energy released during atom-atom collisions in the photoassociation process. Subsequent application of red-detuned light then assures the preparation of maximally one atom in the trap. Using $\Lambda$-enhanced grey molasses for loading, we study and model the molecular processes and find we can trap single atoms with 90$\%$ probability even in a very shallow optical tweezer. Using 100 traps loaded with 80$\%$ probability, we demonstrate one example of the power of enhanced loading by assembling a grid of 36 atoms using only a single move of rows and columns in 2D. Our insight will be key in scaling the number of particles in bottom-up quantum simulation and computation with atoms, or even molecules.

\end{abstract} 


\maketitle 
\normalsize
\section{Introduction}

In quantum simulation and computing, the assembly of large arrays of individually-controllable particles is a frontier challenge. Ultracold gases of neutral atoms have long simulated quantum physics on a macroscopic scale, and quantum gas microscopes are now a window to microscopic dynamics~\cite{bakr_quantum_2009,sherson_single-atom-resolved_2010}.
However, the desire for control of individual atoms, in particular for quantum computing, motivates pursuing bottom-up engineering of neutral atom arrays~\cite{isenhower_demonstration_2010,wilk_entanglement_2010,kaufman_two-particle_2014,kaufman_entangling_2015}.  In a Maxwell's demon approach, experimenters image single atoms and subsequently rearrange them into a desired pattern.  The resulting ordered arrays have presented new opportunities in studies of multi-particle quantum dynamics~\cite{weiss_another_2004,miroshnychenko_quantum_2006,endres_atom-by-atom_2016, barredo_atom_by_atom_2016,bernien_probing_2017,marcuzzi_facilitation_2017,lee_defect-free_2017,barredo_synthetic_2018, kumar_sorting_2018}.  Yet, compared to trapped ions, single neutral atoms are still difficult to trap and assemble.  

\begin{figure}[t]\centering \includegraphics[width=\columnwidth]{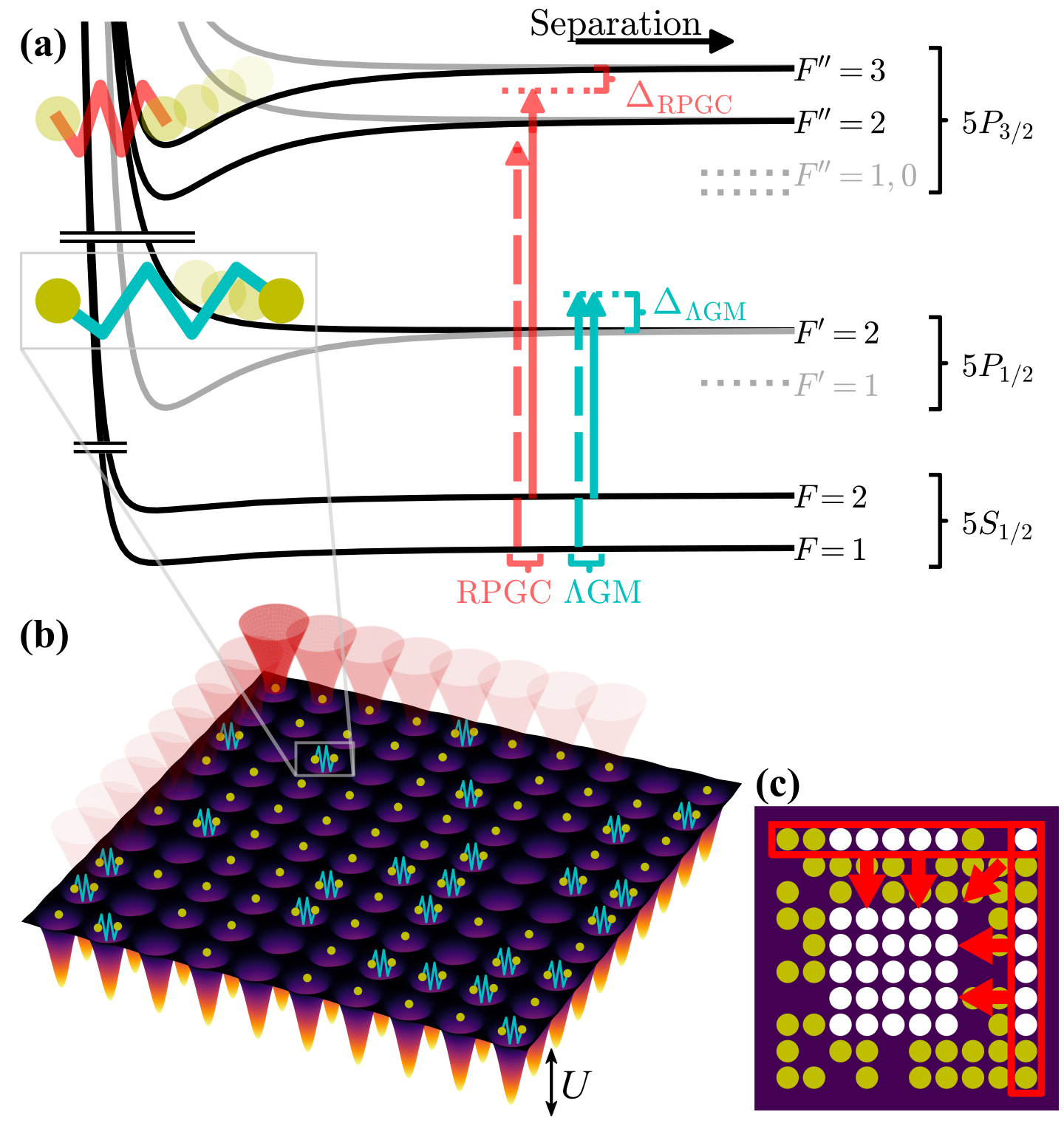}
\caption{
Enhanced loading and rearrangement in large arrays.
\textbf{(a)} Sketch of laser configuration, and molecular energies versus interatomic separation. Solid (dashed) arrows show cooling (repump) lasers with indicated detunings.
\textbf{(b)} Atoms loaded into an array of $100$ traps with depth $U$ formed by optical tweezers undergo blue-detuned light-assisted collisions. \textbf{(c)} Schematic of parallel rearrangement to form defect-free array with target atoms (white) after removing a subset of loaded atoms (yellow).
}
\label{fig:Figure_1}
\end{figure}

In our work, we form ordered atom arrays by combining dense loading of large optical tweezer arrays with atom imaging and rearrangement [Fig.~\ref{fig:Figure_1}].  
Using $\Lambda$-enhanced grey molasses (\LGM{}) on the D$_1$ line of $^{87}$Rb~\cite{Grier_enhanced_2013, RioFernandes_sub_2012}, we can load single atoms with high efficiency in a trap shallower than required for standard sub-poissonian loading~\cite{schlosser_sub-poissonian_2001} and nearly an order of magnitude shallower than required for previous enhanced loading~\cite{grunzweig_near-deterministic_2010}. While we demonstrate the idea with an array of optical tweezers in 2D, dense loading could also be used in optical lattices or in microtraps in 3D~\cite{barredo_synthetic_2018, kumar_sorting_2018}. We predict our technique will scale-up neutral-atom array assembly by expanding rearrangement algorithms and by enabling considerably larger ordered arrays. 

To isolate single atoms in optical tweezers or lattices, one typically drives light-assisted collisions in the collisional blockade regime using red-detuned light~\cite{schlosser_sub-poissonian_2001,schlosser_collisional_2002}. In this case, atoms are photoassociated to attractive molecular states in which they accelerate towards each other and gain kinetic energy that predominantly expels both from the trap [Fig.~\ref{fig:Figure_1}(a)].  If the collisions occur quickly enough to dominate the dynamics, as is the case in microtraps, a single atom is left about half the time. In the pioneering work of Ref.~\cite{grunzweig_near-deterministic_2010}, after adding a blue-detuned laser to drive atoms into repulsive molecular states, the energy gained in the collision was tuned to induce single atom loss~\cite{grunzweig_near-deterministic_2010,carpentier_preparation_2013, lester_rapid_2015,fung_efficient_2015}.
Loading efficiency was enhanced to 90\%, but at the cost of requiring large trap depths ($U/k_B \sim 3$ mK compared to 1 mK for red-detuned loading) and hence making use of the technique untenable in large arrays. 

Here we resolve the conflict that has existed in previous work with enhanced loading -- namely that red-detuned cooling drives lossy collisions and competes with desired blue-detuned collisions. By using \LGM{}, we have the ability to cool into the trap and photoassociate with the same blue-detuned laser [Fig.~\ref{fig:Figure_1}(a)], and we can control the energy atoms are given in the collision by varying the laser's detuning. Further, we can make use of red-detuned and blue-detuned molecular photoassociation processes at will.  In particular, we first modify the atom number distribution in the microtrap with blue-detuned cooling (\LGM{}).  We then apply red-detuned light, which both assures that not more than a single atom remains and, if it remains, images it.  The loading behavior studied in a single trap agrees with a model of consecutive light-assisted collisions to repulsive molecular states. Our model further allows us to identify paths to even more efficient single-atom loading. 

We find we can load a single optical tweezer with a trap depth of $U/k_B = 0.63(6)$~mK with 89(1)\% efficiency, and a $10\times10$ array with 80.49(6)\% efficiency [Fig.~\ref{fig:Figure_1}(b)].  We also demonstrate a proof-of-principle rearrangement technique that relies on the enhanced loading to create a $6\times6$ defect-free array using a simplified sequence of parallel moves of entire rows and columns [Fig.~\ref{fig:Figure_1}(c)]~\cite{endres_atom-by-atom_2016}.  Lastly, we discuss how the efficiency of both this simplified rearrangement, as well as atom-by-atom assembly, scale exponentially with initial filling of the array.

\section{Loading Studies and Modeling}

\subsection{Loading Experiments}

Generally, in $\Lambda$-enhanced grey molasses (\LGM{})~\cite{RioFernandes_sub_2012,Grier_enhanced_2013} the cooling laser is set blue-detuned of a type-II ($F'\le F$) transition and in a $\Lambda$ configuration with a coherent repump laser  [Fig.~\ref{fig:Figure_1}(a)]. Because of its greater isolation from nearby hyperfine manifolds, we chose to operate the \LGM{} detuned from the $5\text{P}_{1/2}\ket{F'=2}$ state [in contrast to, e.g.~$5\text{P}_{3/2}\ket{F''=2}$]. 
Note, we were motivated to use \LGM{} mainly as a natural way to blue-detune both cooling and repump lasers, which is a somewhat different motivation than in recent quantum degenerate gas experiments with light atoms and molecules -- namely that grey molasses works on open transitions, and $\Lambda$-enhancement results in lower temperatures~\cite{aspect_laser_1988,shahriar_continuous_1993,Grynberg_proposal_1994,Esslinger_purely_1996,RioFernandes_sub_2012,Grier_enhanced_2013,mccarron_improved_2015,devlin_three-dimensional_2016,cheuk_lambda-enhanced_2018,lim_laser_2018,anderegg_laser_2018,Rosi_enhanced_2018,bruce_sub-doppler_2017}. 

\begin{figure*}[!t]\centering
\includegraphics[width=\textwidth]{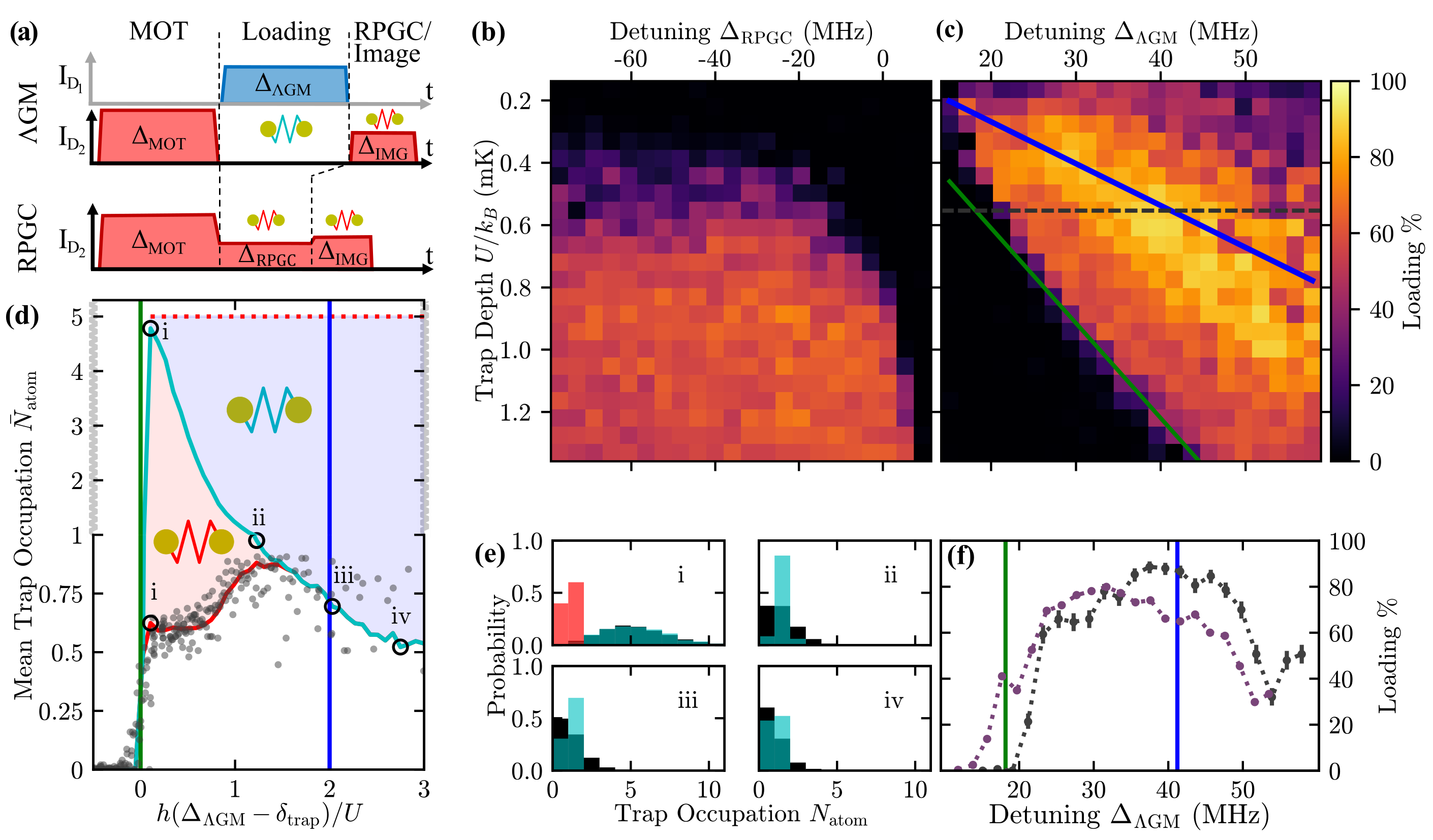}
\caption{\LGM{} comparative loading studies. \textbf{(a)} Flow-diagram of experiments. \textbf{(b,c)} Average loading efficiency in to a single optical tweezer from $150$ experiment repetitions as a function of bare laser detuning from the free-space resonance and trap-depth, for RPGC (b) and \LGM{} (c).   \textbf{(d)} Monte-Carlo simulated mean trap occupation with bi-linear vertical scale (black and grey axes). In the \LGM-step, the initial trap occupation (dashed, red line) is reduced (cyan line) by blue-detuned collisions (blue area). In the RPGC imaging step, red collisions (red area) further reduce the trap occupation (red line).  The resultant is compared with data (grey points) from (c) averaged for $U/k_B>0.65~$mK (see text). \textbf{(e)} Histograms (cyan) of trap occupancy from the Monte-Carlo after the \LGM-step for detunings indicated by the black circles in (d), compared to a Poisson-distribution (black) with the same mean trap occupation, and atom distribution after the RPGC-step (red in i). \textbf{(f)} Loading efficiency as a function of \LGM{}-laser detuning $\Delta_\text{LGM}$ for a single tweezer (black) [see also cut along the black dashed line in (c)] and a regular array of $100$ tweezers (purple) at $U/k_B\approx0.55(5)~$mK. Error bars indicate statistical $1\sigma$-confidence interval (see Appendix).  Throughout panels, green lines are AC-Stark-shifted $\ket{F=2}-\ket{F''=3}$ transition (for RPGC) and AC-Stark-shifted $\ket{F=2}-\ket{F'=2}$ transition $\delta_{\mathrm{trap}}$ (for \LGM), and blue lines are $\delta_{\mathrm{trap}}+2U/h$.
}
\label{fig:Figure_2}
\end{figure*}

We first present results from loading a single optical tweezer using \LGM{}, and compare to standard loading using red-detuned polarization gradient cooling (RPGC) [Fig.~\ref{fig:Figure_2}(b,c)]. We capture $^{87}$Rb atoms in a magneto-optical trap (MOT) and then cool them into a spatially-overlapped optical tweezer with depth $U$ with either \LGM{} or RPGC. After the cooling and loading stage, we apply RPGC with parameters optimized for fluorescence imaging of the atoms.  Initially, this quickly removes remaining atom pairs, and then images whether a single atom or no atom remains in the trap [Fig.~\ref{fig:Figure_2}(a)]. The procedure is repeated to determine average single-atom loading efficiencies, i.e., the fraction with which a single atom is found after both the loading and imaging stages.  See Fig.~\ref{fig:Figure_2}(a) and the Appendix for experimental-sequence timing and details of the imaging analysis. Also, see Fig.~\ref{fig:Figure_1}(a) and the Appendix for detailed laser configurations.  

Fig.~\mbox{\ref{fig:Figure_2}}(b,c) shows the loading probability $P$ as a function of both laser detuning from the closest atomic free-space resonance and trap depth, for both RPGC and \LGM{}.  With \LGM{} we observe $89(1)\%$ loading efficiency at $(\DLGMM,~U/k_B)=(45~\text{MHz},~0.55(5)~\text{mK})$, and we can still load with $\sim 80\%$ efficiency at trap depths of $U/k_B\approx 0.27(3)~$mK.  These findings are remarkable as with the same optical power we can load tweezer arrays that are more densely filled \emph{and} two to three times larger compared to RPGC loading.
The maximum RPGC loading of $64(1)\%$ for $(\Delta_\text{RPGC},~U/k_B)\approx(-14~\text{MHz},~1.1(1)~\text{mK})$ is among the highest reported for RPGC~\cite{lester_rapid_2015,fung_efficient_2015,barredo_atom_by_atom_2016,endres_atom-by-atom_2016}.  In the simplest picture of RPGC, one expects $50\%$ loading, but, in agreement with other studies~\cite{fung_efficient_2015}, additional processes result in $\sim35\%$ of the collisions causing only one atom to leave the trap.

A physically rich picture can be gained from studying the detuning dependence of \LGM{} loading [Fig.~\ref{fig:Figure_2}(c)]. First, note that the trap light results in an AC Stark shift $\delta_\text{trap}=32.8 \tfrac{\text{MHz}}{\text{mK}}\tfrac{U}{k_B}$ of the atomic transition in the center of the trap [green lines in Fig.~\ref{fig:Figure_2}(c,d,f)]. The blue line in Fig.~\ref{fig:Figure_2}(c,d,f), which marks a shift of $2 U/h$ from the trap-shifted resonance, is a key energy scale for the physics of the enhanced loading. At shifts smaller than $2 U/h$, the collision does not give a pair of zero-temperature atoms sitting at the bottom of the trap enough energy for either to escape, while at larger detunings both atoms will be expelled. A finite temperature, and hence an initial center of mass motion, will blur the transition, and indeed is necessary for inducing the desired single-atom loss. Although our data are roughly consistent with this picture, we look more closely by plotting the data of Fig.~\ref{fig:Figure_2}(c) against a dimensionless detuning $h(\DLGMM - \delta_{\text{trap}})/U$.
We do this for all data traces $U/k_B\geq0.65~$mK [Fig.~\ref{fig:Figure_2}(d)], and observe a number of interesting features. For example, we observe a $\sim60\%$ loading probability for small detunings and that the maximum loading peaks below the $2 U/h$ shift (blue line).

\subsection{Model}

To elucidate detailed trends, we have carried out a Monte-Carlo calculation of the collision dynamics. Most generally, we expect loading to be affected by both collisions and the \LGM{} cooling performance, and both may be influenced by the non-trivial light shifts and polarization gradients in the tweezer traps.  Modeling the interplay of these effects is beyond the scope of this work, but we can understand the collisional process quantitatively if we assume the continuous \LGM{} cooling can load at least a few atoms per trap, and re-thermalizes any atoms remaining after a collision. The simulation starts by preparing a Poisson-distributed number of atoms $N_\text{atom}$ with a mean number $\bar{N}_{\rm{atom}}=5$ and temperature $T$, where $\bar{N}_{\rm{atom}}$ was chosen $>2.5$ to avoid loading zero atoms initially. To simulate the finite experiment cooling time, we calculate a finite number of 5000 time steps each having two atoms collide once if they are closer than 100 nm. A collision might eject none, one, or both atoms out of the trap depending on the final energy of each atom, which is determined by their pre-collision energy and the collisional energy gain $E=h\left[\DLGMM-\delta_{\rm{trap}}\right]$. This process continually reduces $N_{\rm{atom}}$ in each time step. At the end, the RPGC imaging is simulated by assuming that it entails a fast collisional process at the start of the image where red-detuned collisions reduce atom numbers in a manner consistent with our red loading - namely we reduce any remaining $N_\text{atom}>1$ by 2 with a chance of 65\% and by 1 with a chance of 35\% until $N_\text{atom}\leq1$. 
 
Fig.~\ref{fig:Figure_2}(d) shows the result of the Monte-Carlo simulation by indicating the mean trap occupation $\bar{N}_{\rm{atom}}$ as a function of the normalized collisional energy gain. During \LGM{}-loading, the initial atom number (red dashed line) is reduced (cyan line). During RPGC imaging $\bar{N}_{\rm{atom}}$ is further reduced [red line in Fig.~\ref{fig:Figure_2}(d)]. Fig.~\ref{fig:Figure_2}(e) shows the simulated atom-number distribution ($N_{\rm{atom}}$) in the trap and how atom loss in the \LGM{} and RPGC-phase modifies the Poisson distribution. 

We observe three physical regimes: For $E\ll2U$ [see panel i in Figure 2(e)], the \LGM{}-phase has little effect as almost no atom loss occurs, hence the initial poisson-distribution (black distribution) is not modified (blue distribution). The initial phase of the RPGC-imaging step, however, reduces the number of atoms to 0 or 1, yielding a RPGC-like 65\% mean trap occupation (red distribution). In contrast, for $E\gg2U$ (panel iv), two-body losses dominate in the \LGM-phase as every collision expels both atoms from the trap resulting in $N_\text{atom}=0$ ($N_\text{atom}=1$) in $\sim50\%$ of the cases if the loaded $N_\text{atom}$ was even (odd). Hence after \LGM, $N_\text{atom}<2$ and the red-detuned imaging phase does not modify the atom-number distribution anymore. At the transition $E\approx 2U$ (panel iii), both single atom and two-body losses occur in the \LGM-phase with roughly equal probability because of the finite temperature. Since two-body losses tend toward an equal distribution of $N_\text{atom}=0$ and $1$, and single-atom loss toward $N_\text{atom}=1$, we load a single atom ($N_\text{atom}=1$) in 75\% of the cases. Again, RPGC-imaging does not modify the distribution as $N_\text{atom}<2$ after \LGM. Maximal loading probability is found at $E<2U$ (panel ii) where only single-atom loss occurs. Here, any occurrence of $N_\text{atom}=0$ is a result of either no atoms having been loaded initially, or the \LGM-step has not finished (finite $N_\text{atom}>1$ after \LGM), and RPGC-imaging then ejects pairs atoms.

Our model indicates no fundamental limitations to the loading efficiency and that by optimizing the trap size, atom temperature, and related parameters, it may be possible to reach higher loading fractions. Note that the only free parameter that affects the prediction of the simulation is the atom temperature $T$ in the trap. The simulation describes our data well for $T=120(10)~\text{\micro K}$, which needs to be understood as an average value for the different trap-depths $U$ that were investigated. This value is close to the free-space \LGM{} temperature we measure of $T\approx50~\text{\micro K}$, which is higher than typical values, likely due to non-ideal beam geometries (see Appendix).

\section{Impact of Grey Molasses Loading on Array Assembly}

\subsection{Loading in Large Arrays}

We have also performed a loading study for an array of $10\times10$ optical tweezers spaced by 2 $\mu$m.  We display the measurement at $U/k_B=0.55(5)~$mK  as the purple line in Fig.~\mbox{\ref{fig:Figure_2}}(f). 
Compared to the single-trap data at similar $U$ (black), the data are shifted to smaller detunings, and we observe a maximum loading of $80.49(6)\%$ in a single run averaged over the $10\times10$ array. 
These effects could be due to a variety of consequences of the larger array: variations in trap shape and depth or overall degradation of the optical spot sizes (see Appendix).  Note that our experimental apparatus was designed and optimized to entangle closely-spaced atoms in ground states, in contrast to systems that interface (farther-spaced) Rydberg atoms. This places constraints on the tweezer array, such as acousto-optic device mode, trap-light detuning, optical power, and high-NA field-of-view, that mean with typical loading we are limited to working with array sizes less than $10\times10$.  However, \LGM-loading allows us to scale up both the size of our arrays as well as the total number of atoms we can trap, and is a unique realization for enhanced loading in optical tweezers. 

\begin{figure}[t] \centering
\includegraphics[width=\columnwidth]{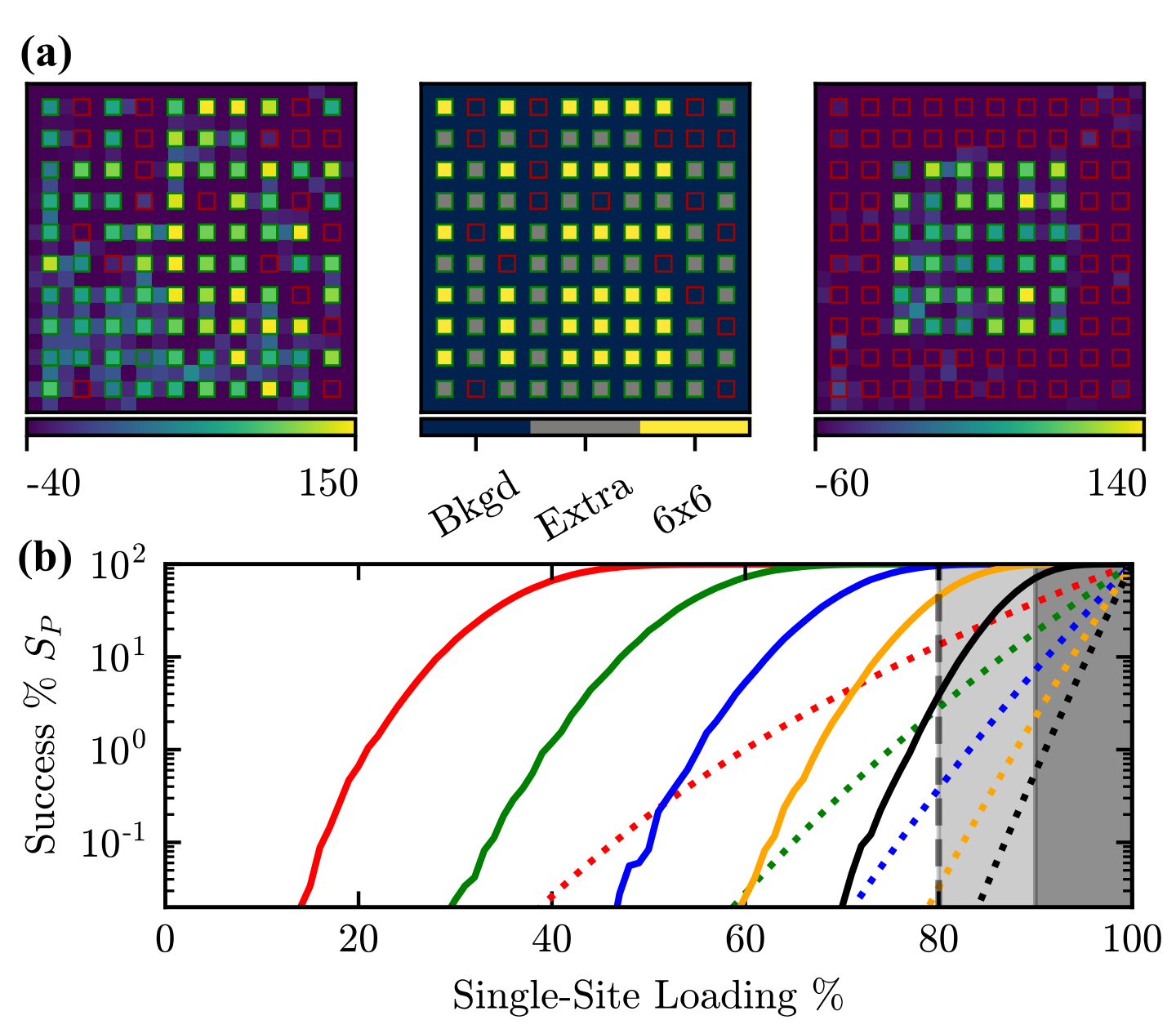}
\caption{Parallel rearrangement in 2D. \textbf{(a)} Single image of 80 atoms in an array of 100 traps with separation 2 $\mu$m (left). Binarized loading (center) indicating empty traps (red squares), occupied traps (grey pixels), and loaded traps selected for parallel rearranging (yellow pixels) to a defect-free $6\times6$ array. For the single-shot images, in each trap pixel (background pixel) the atom count threshold (average threshold of nearest neighbors) is subtracted for clarity from the count number recorded by the camera (see Appendix). \textbf{(b)} Monte-Carlo-simulated success probabilities ($S_P$) of finding a filled disjoint array of atoms within $10\times10$ traps as a function of loading efficiencies (solid lines), and corresponding loading probability of a specific array without rearrangement (dashed lines). Colors indicate target array sizes: $3\times 3$ (red), $4\times 4$ (green), $5\times 5$ (blue), $6\times 6$ (orange), and $7\times 7$ (black). The light grey area is only accessible with average loading greater than $80$\%, and the dark grey area for greater than $90$\%.
}
\label{fig:Rerng_simulations}
\end{figure}

\subsection{Rearrangement Example}

One prospect for dense loading is that it opens up new possibilities for rearrangement algorithms for array assembly.  Here we present one particular experimental example.  After densely loading an array, we first obtain the location of each atom using a single image [left panel Fig.~\ref{fig:Rerng_simulations}(a)].
Even with dense loading, the probability of loading a specific set of $6\times6$ traps is exponentially small~[dashed lines in Fig. \ref{fig:Rerng_simulations}(b)]. 
However, there are many potential sets of (sometimes disjoint) $6\times6$ traps embedded in the $10\times10$ array. 
We then search for such a configuration of completely loaded $6\times 6$ traps.
If successful, we turn off the extra traps to remove the excess atoms, and then contract and shift the identified disjoint array in a \emph{single} move (right panel)~\cite{endres_atom-by-atom_2016}.
Currently, successful rearrangement to a square $n=36$ (6 by 6) array only works in $0.1\%$ of cases due to  unexpected loss observed when turning off rows and columns, in which an atom is effectively lost with a $17\%$-chance. This loss is technical in nature, and potential causes include intermodulation in the tweezer-generating rf tones or collisions between the trapped and dropped atoms.  But, as illustrated in Fig.~\mbox{\ref{fig:Rerng_simulations}}(b), observing this array with this parallel technique would have been impossible without enhanced loading.
In increasing the loading probability $P$ from $60\%$ to $80\%$, the percentage of experiment runs in which one could possibly extract a defect-free $6\times6$ array goes from $0.02\%$ to $37\%$.
Notably, this entire procedure is completed using only a pair of acousto-optic modulators to control the optical tweezers.

\subsection{Scaling Arguments}

 The full potential of dense loading using \LGM{} will come when combined with the most advanced atom-by-atom rearrangement algorithms in 2D or 3D~\cite{barredo_atom_by_atom_2016, barredo_synthetic_2018,kumar_sorting_2018}. In principle, if the Maxwell's demon atom rearrangement operation is perfect, and if a large number of traps are used, one would expect very large arrays could be created through a bottom-up approach irregardless of the loading.  Practically, however, there are many factors that steeply limit creating large-scale systems with optical tweezer assembly - in particular, a finite atom lifetime compared to the time required for rearrangement, and also simply the number of traps that can be created and loaded.
 
 Our approach addresses these problems because it is efficient in shallow traps.  Hence, with \LGM{} more loadable traps can be created with the same amount of optical power. Additionally, 2D algorithms fill defects in a target array of size $n$ with a sequence of $m\propto (1-P)n^{1.4}$ single moves, which we verified with Monte-Carlo simulations~\cite{barredo_atom_by_atom_2016}. In scaling up array sizes, the time and number of moves required become lengthy, lowering the probability of successful rearrangement ($S_P$) as errors $\epsilon$ due to finite move fidelities and background collisions suppress this success rate as $e^{-m\epsilon}$ \cite{barredo_atom_by_atom_2016}. Increasing the loading probability $P$ from $P=60\%$ to $P=90\%$ would decrease $m$ by a factor of $4$, making larger array sizes more obtainable and exponentially improving the success probability $S_P$.  

\section{Conclusion}

In conclusion, by gaining control over photoassociation to molecular states we have demonstrated enhanced loading of arrays of shallow optical tweezers. As described in Secs III A and B, we achieved a strong relative improvement on our trapped atom-numbers, but experimental platforms designed to host more optical traps than our system will stand to benefit even more.  For example, Ref.~\cite{barredo_atom_by_atom_2016} loads approximately 50 atoms with a 2D-array of 100 traps of 1 mK depth. With the same optical power and \LGM, one would expect to utilize 370 traps of 0.27 mK depth and, based on the shallow-depth loading of single atoms at $P=80\%$ of Fig.~2(c), load approximately 300 atoms -- a six-fold increase.  Further, the density of the filling will affect the number of moves required in rearrangement~\cite{endres_atom-by-atom_2016}.  Using a technique that moves atoms individually~\cite{barredo_atom_by_atom_2016}, our Monte-Carlo simulations indicate that rearranging 300 atoms at $P=50\%$ requires approximately $900$ moves on average, whereas at $P=80\%$ it requires 320 moves. As a result, the probability to retain all 300 atoms in the rearrangement protocol increases roughly from $0.1\%$ to $10\%$ when going from $P=50\%$ to $P=80\%$, assuming a 420 second atom lifetime~\cite{covey_2000-times_2018}, 1 ms per move, and a $99.3\%$ move fidelity~\cite{barredo_atom_by_atom_2016}.
  
While we have studied one particular blue-detuned cooling mechanism -- \LGM{} on the $^{87}$Rb $5S_{1/2}-5P_{1/2}$ transition -- it will be interesting to explore a variety of other related cooling techniques in future experiments.  In particular, it is also known that grey molasses is effective on the $5S_{1/2}-5P_{3/2}$ transition~\cite{Rosi_enhanced_2018}, and future studies could compare the salient molecular physics in each manifold~\cite{weiner_experiments_1999}.  Further, we expect our work will be the start of explorations of the interplay of collisions and cooling in microtraps for a host of blue-detuned cooling mechanisms with alkali atoms, other atomic species, and even molecules~\cite{liu_building_2018,cheuk_lambda-enhanced_2018}.

\section*{Acknowledgements}

We thank Yiheng Lin, Kai-Niklas Schymik, Ludovic Brossard, Junling Long, and Brian Lester for technical assistance, and Jose D'Incao, Paul Julienne, Ana Maria Rey, and Adam Kaufman for helpful comments and fruitful discussion.  We acknowledge funding from ONR Grant No. N00014-17-1-2245, NSF Grant No. PHYS 1734006, the Cottrell Scholars program, and the David and Lucile Packard Foundation.  M. O. B. acknowledges support from an NDSEG Fellowship.  T.T. acknowledges funding from SNF under project number: P2EZP2\_172208.

\section*{Appendix}

\subsection*{1. Optical Tweezers}
We generate an array of optical-tweezer traps spaced by $2~\text{\micro m}$ in the $xy$-plane by passing a single 850~nm laser beam through two orthogonal longitudinal-wave $\mathrm{TeO_2}$ acousto-optical modulators (AOMs) with center frequencies (bandwidths) of 180~MHz (90~MHz).
Each modulator is driven with a sum of radio-frequency (RF) tones with frequency (amplitude) that can be individually and dynamically adjusted to control the position (intensity) of different tweezer-rows and columns. The relative phases of the tones are set to minimize intermodulation in the RF setup. The array of deflections created by the AOMs is then imaged by a 0.6-NA-objective lens into a glass cell. This creates a trap with a $0.68~\text{\micro m}$ waist for a single tweezer, and traps with an average waist of $0.75~\text{\micro m}$ for a $10\times 10$ array. The standard deviation of the trap depths was minimized to 8\% by optimizing the RF amplitudes. Trap depths are calibrated by measuring light-shifts of in-trap atomic transitions as a function of trap power and applying a linear fit; the slope gives a calibration of the intensity of trap light the atom experiences, which can be be used to directly calculate the trap depth~\cite{kien_dynamical_2013}. Errors on trap depths are 1-$\sigma$ errors extrapolated from the errors on the slope of the linear fit. The lifetime of atoms in the traps is limited to 5~sec by the background pressure.

\subsection*{2. Laser Cooling and Loading}

In all experiments, three beam paths are used to address the atoms. Two (diagonal) paths are along the diagonals of the $xy$-plane, and a third (acute) path in the $xz$-plane is at an angle of 55\degree{} from the z-axis to avoid the objective~\cite{kaufman_laser-cooling_2015}. All lasers along these paths are retro-reflected and in a $\sigma^+\sigma^-$ polarization configuration.

Our magneto-optical trap (MOT) is spatially overlapped with the trap array and cools atoms for 500 ms to a temperature of $\sim 100~\text{\micro K}$, measured by imaging its ballistic expansion. The cooling (repump) laser is red-detuned from the $D_2$  $\ket{F=2}\rightarrow\ket{F''=3}$ ($\ket{F=1}\rightarrow\ket{F''=2}$) transition, and applied on all three beam paths (on only the diagonal paths). In the case of the 20-ms-long RPGC stage we cool the atoms to $\sim 10~\text{\micro K}$. For this, we detune the cooling (repump) laser by $\Delta_\text{RPGC}$ (20~MHz), set the intensities at $1.3~I_\text{sat}~ (0.1~I_\text{sat})$ on the diagonal paths and $4.5~I_\text{sat}$ ($0~I_\text{sat}$) on the acute path, and zero the magnetic fields. 

In the case of the 200-ms-long $\Lambda$-enhanced grey molasses (\LGM{}) stage, we apply a cooling laser that is detuned by \DLGM{} from the $\text{D}_1$ $\ket{F=2}\rightarrow \ket{F'=2}$ transition at $2.5~I_\text{sat}$ ($0.4~I_\text{sat}$) on the acute (diagonal) paths. We create the coherent repump beam from the cooling laser on the acute path using an electro-optic modulator. The repump beam is detuned by $\DLGMM+0.14~$MHz from the $\text{D}_1$ $\ket{F=1}\rightarrow \ket{F'=2}$ transition and at $1.5~I_\text{sat}$. Note that the optimal \LGM{} free-space temperature of $50~\text{\micro K}$ is reached for $\DLGMM\approx15~\text{MHz}$ and is likely limited by the beam path geometry and repump light configuration.

\subsection*{3. Imaging, Data, and Statistics}

Regardless of the loading configuration, we image the atoms using another RPGC stage with the cooling beam $\Delta_\text{RPGC}=-19~$MHz at $3~I_\text{sat}$) only on the acute path. We alternate the tweezer-light with the imaging light at 2~MHz to scatter light when atoms are experiencing no light shifts. This configuration is maintained for 20~ms during which we collect scattered photons on an EMCCD camera, superbinned to $4\times4$ pixels to reduce readout noise.
As we now discuss experimental evidence for, this red-detuned imaging process quickly kicks out any pairs of atoms that might exist, for example, in the case of a grey molasses loading stage with a small detuning [Fig.~\ref{fig:Figure_2}(d) and (e)]. Accordingly, it does not resolve a tweezer's occupation number following the grey molasses stage [illustrated by the cyan line in Fig.~\ref{fig:Figure_2}(d)], but rather maps the atom number onto 0 or 1.  If this loss did not occur quickly compared to the imaging time, we would sometimes collect numbers of photons significantly larger than our calibrated single-atom scattering rate. We do not observe this signature despite high experimental statistics, suggesting a sub-percent impact of these effects on the imaging. 

At every atom location individually, to determine a count threshold that indicates the presence of an atom in the trap, we create a histogram of all counts during an experiment and fit it with a sum of two Gaussians.  The threshold with maximal fidelity $\mathcal{F}$ is found, where $\mathcal{F}=1 - (E_{fp} + E_{fn})$, with $E_{fp}$ ($E_{fn}$) being the expected rate of false positives (false negatives) from the fits. This converts a sequence of counts to a sequence of Booleans which is averaged to determine the loading probability. By finding thresholds for each trap individually and subtracting them from the images in Fig.~\mbox{\ref{fig:Rerng_simulations}}(a), we compensate for a spatially varying background noise and, due to the limited field of view of our high-NA lens, the different numbers of photons we collect for each trap.

All errors reported indicate 1-$\sigma$ equal-tailed Jeffrey's prior confidence intervals~\cite{brown_interval_2001}. The loading efficiencies reported in the main text for RPGC (64(1)\%), \LGM{} (89(1)\%), and $10\times10$-\LGM{} ($80.49(6)\%$) were obtained by analyzing 2000, 1000, and 5000-per-atom repetitions with threshold fidelities 0.987, 0.998, and 0.993 respectively.


%

\end{document}